\documentclass[sigconf,nonacm]{acmart}
\AtBeginDocument{%
  }

\usepackage{mathtools}
\usepackage{subcaption} 
\usepackage{soul}
\usepackage{verbatim}
\usepackage{hyperref}
\hypersetup{
    colorlinks=true,
    linkcolor=blue,
    filecolor=magenta,      
    urlcolor=cyan,
    pdftitle={TRACE: Traversal and Reasoning Algebraic Computing Engine for Formal Hardware Verification}
    }

\usepackage{subcaption}

\usepackage{booktabs}
\usepackage{multicol}
\usepackage{multirow}
\usepackage{tabularx}

\usepackage{enumitem}
\begin{document}

\title{TRACE: Traversal and Reasoning Algebraic Computing Engine for Formal Hardware Verification}


\author{Jan Kleinekathöfer, Lennart Weingarten}
\authornote{Both authors contributed equally to this research.}
\email{{ja\_kl, len\_wei}@uni-bremen.de}
\affiliation{%
  \institution{Institute of Computer Science, University of Bremen}
  \city{Bremen}
  \state{Bremen}
  \country{Germany}
}

\author{Kamalika Datta, Rolf Drechsler}
\email{{kdatta,drechsler}@uni-bremen.de}
\affiliation{%
  \institution{Institute of Computer Science, University of Bremen/DFKI}
  \city{Bremen}
  \state{Bremen}
  \country{Germany}
}

\renewcommand{\shortauthors}{Anonymous et al.}

\begin{abstract}


Modern hardware verification of complex circuits relies heavily on the efficiency of formal methods.
For complex arithmetic circuits in particular \emph{Symbolic Computer Algebra} (SCA) engines which represent pseudo-boolean functions using polynomials are crucial. As circuit complexity grows in the age of AI, verification of arithmetic primitives, including Multiplication, Addition, \emph{Multiply-Accumulate} (MAC), becomes a computational bottleneck. To address this, we introduce TRACE (Traversal and Reasoning Algebraic Computing Engine), a highly efficient framework designed to investigate the intersection of traversal strategies and proof efficiency.

Unlike existing SCA tools which are mainly limited to multipliers, TRACE offers a flexible framework for researchers to analyze memory usage and verification time across a wide range of arithmetic circuits (adder, multiplier and MAC). To overcome the state-explosion problem inherent in polynomial expansion, the engine incorporates advanced reduction techniques, including optimized traversal strategies, conflict removal, and polarity-based optimization for compact symbolic representations. Our experimental results show that for optimized MAC, for the first time, TRACE was able to verify previously unverifiable circuits, achieving an overall speedup of 410x in verification time.


\end{abstract}



\maketitle

\section{Introduction}\label{sec:introduction}

The verification of modern integrated circuits increasingly depends on the mathematical correctness provided by formal methods. From AI-accelerated tensor cores to high-fidelity digital signal processors, the demand for complex structures like multiplication, addition, \emph{Multiply-Accumulate} (MAC), has never been higher. However, as these circuits grow in bit-width and architectural complexity, ensuring their functional correctness becomes a critical challenge. Although other formal proof engines based on \emph{Binary Decision Diagrams} (BDD)~\cite{B:1986}, \emph{Satisfiability Solvers} (SAT)~\cite{moskewicz2001chaff}, or \emph{Answer Set Programming} (ASP) exist, \emph{Symbolic Computer Algebra} (SCA)~\cite{kaufmann2019verifying,KBK:2019,MGD:2022_RevSCA2,konrad2024FMCAD,Fujita24,HPJHMHYB:2024} engines have become the standard for ensuring the functional correctness of multiplier and MAC designs. This is performed by representing circuit logic as a system of algebraic polynomials. This method allows for a proof of correctness by checking if the implementation design matches the specification.

SCA has shown promising results in the formal verification of simple multipliers. Following this initial success, more complex multipliers and related circuit types have been studied. RevSCA~\cite{MGD:2022_RevSCA2}, AMulet~\cite{kaufmann2021amulet} and DynPhaseOrderOpt~\cite{konrad2024FMCAD} are the most popular SCA engines showcasing the efficiency of SCA. Although none of these tools currently support MAC verification. In recent times, structurally known MAC and DP circuits are also verified using the extensions of RevSCA~\cite{WDD:2025:fdl,WDD:2025:date}, although they fail to verify structurally unknown or complex architectures even for bit size greater than 11~\cite{WDD:2025:dvcon}. A persistent challenge in the verification of complex circuits is the intermediate term explosion during polynomial reduction. To mitigate this, various optimizations have been introduced, which can be broadly categorized into two groups: 
(1) \textit{Ordering optimization} techniques aim to identify the optimal substitution order to minimize the size of intermediate polynomials.
(2) \textit{Representation optimizations} try to reduce the representation size of a specific intermediate polynomial. 
This raises the question: what are the effects of different ordering strategies (both static and dynamic) and representations on various circuit types and architectures? Specifically, is there a particular ordering or representation that is optimally suited for verifying a specific class of circuits, or can a generalized method be developed that remains effective across all circuit types? This leads to a large configuration space of circuit types and optimizations. So far, no SCA-based tools are available that can allow for such an exploration. Existing tools either support only multiplier verification or fail to verify complex MAC circuits. To enable a systematic examination of the configuration space, we introduce TRACE. TRACE facilitates the combination of traversal strategies and optimization techniques across diverse circuit architectures without compromising on computational performance which existing tools lack. Our tool can be accessed using this link \footnote{\url{https://github.com/jan-kl/trace}}.   
Experimental results reveal that for optimized MAC circuits, overall speedup of 410x in verification time is achieved. It also provides interesting findings regarding how this tool can be utilized to select the optimal technique for a particular circuit type.
Listed below are the main features of TRACE:
\begin{enumerate}
    \item Implementation of adapted \emph{Specification Polynomial} (SP) for adder, multiplier and MAC designs.
    \item Incorporation of various static and dynamic ordering techniques for traversal.
    \item Integration of advanced reduction technique like polarity optimization and conflict removal.
    \item Providing reports on a range of relevant metrics and offering debugging output for various stages of the verification flow to provide insights into the verification process. 
\end{enumerate}

The remainder of this paper is organized as follows: Section~\ref{sec:background} provides a brief background on SCA. Section~\ref{sec:proposed} details the overall methodology and architecture of TRACE. Section~\ref{sec:engine_capabilities} describes the traversal techniques, symbolic representations, and optimizations employed, followed by the experimental evaluation in Section~\ref{sec:experimental}. Finally, Section~\ref{sec:conclusion} concludes the paper.

\section{Background}\label{sec:background}
The SCA-based verification methodology follows a three-step process: defining the \emph{Specification Polynomial} (SP), identifying \emph{Gate/Node Polynomials} (GP/NP), and performing backward rewriting. Compatible with both gate-level netlists and \emph{And-Inverter Graphs} (AIGs)~\cite{biere2007aiger}, the process begins by modeling circuit functionality as an SP. It then assigns a polynomial (GP or NP) to every internal node. During the core verification phase, the SP is iteratively updated via backward rewriting—substituting node variables in reverse topological order until the primary inputs are reached. A final remainder of zero confirms functional correctness, while a non-zero result indicates a design fault. In this work, the initial circuit is transformed into AIG and then the verification is performed.

For an AIG the following cases for a primary output edge of an AND-gate node can occur, they are summarized as node polynomial rules:
\begin{enumerate}
    \item[r1]: $z = a$; an edge can act as a buffer or simple wire
    \item[r2]: $z = 1 - a$; an edge can act as a complemented edge
    \item[r3]: $z = ab$; AND-gate without any complemented inputs
    \item[r4]: $z = b -ab$; AND-gate with one complemented input edge (\textit{a} complemented) or $z = a -ab$ (\textit{b} complemented)
    \item[r5]: $z = 1 -a -b +ab$; an AND-gate with both complemented input edges
\end{enumerate}

\begin{figure}[h!]
    \centering
    \includegraphics[width=0.8\linewidth]{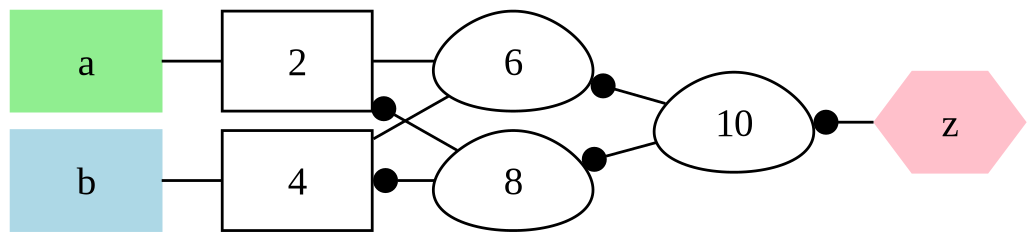}
    \caption{AIG  of the XNOR function z = XNOR (a, b)}
    \label{fig:xnor_aig}
\end{figure}

Consider an XNOR gate with inputs $a, b$ and output $z$ (AIG representation in Fig.~\ref{fig:xnor_aig}). The specification polynomial is as follows:
\begin{equation*}
    SP_{XNOR} = z -1 +a +b +2ab = 0     
\end{equation*}

Fig.~\ref{fig:XNOR_substitution} shows the substitution process, starting with the substitution of the primary output z with rule r2, then continuing with node n10 and n8, which are substituted with r5 and finally substituting node n6 with r3, which results in the zero polynomial, indicating this circuit is correct. For this, the initial polynomial size consist of 5 monomials and the maximum polynomial size goes up to 7 monomials. The maximum polynomial size corresponds to the maximum number of monomials, in a polynomial, in its fully flattened and reduced form~\cite{MGD:2022_RevSCA2}.

\begin{figure}[ht]
{\scriptsize
\begin{align*}
SP_{0}(z, r2) &= z - 1 + a + b - 2ab \\
              &= 1 - n_{10} - 1 + a + b - 2ab \\
              &= -n_{10} + a + b - 2ab \\[1ex]
SP_{1}(n_{10}, r5) &= -(1 - n_6 - n_8 + n_6n_8) + a + b - 2ab \\
                   &= -1 + n_6 + n_8 - n_6n_8 + a + b - 2ab \\[1ex]
SP_{2}(n_8, r5)    &= \begin{multlined}[t] -1 + n_6 + (1 - a - b + ab) \\ 
                      - n_6(1 - a - b + ab) + a + b - 2ab \end{multlined} \\
                   &= \begin{multlined}[t] -1 + n_6 + 1 - a - b + ab - n_6 \\ 
                      + an_6 + bn_6 - abn_6 + a + b - 2ab \end{multlined} \\
                   &= an_6 + bn_6 - abn_6 - ab \\[1ex]
SP_{3}(n_6, r3)    &= a(ab) + b(ab) - ab(ab) - ab \\
                   &= ab + ab - ab - ab = 0
\end{align*}}
\caption{\normalsize Substitution Steps for a XNOR function}
\label{fig:XNOR_substitution}
\end{figure}

\subsection{Related Work}
SCA has emerged as a robust framework for verifying complex arithmetic designs, including multipliers, dividers, and MAC units~\cite{FA:2015, YBL+:2016, SGK+:2016, RBK:2017, YCM:2017, RBK:2018, kaufmann2019verifying, KBK:2019, Fujita24, HPJHMHYB:2024, konrad2024FMCAD, WDD:2025:date}. To address the polynomial explosion in SCA, researchers have developed various mitigation strategies. These include redundant term elimination, algebraic simplification, and column-wise decomposition to partition tasks into sub-problems \cite{KBK:2019}. While reverse engineering helps extract structural relationships for optimized designs \cite{MGD:2022_RevSCA2}, it remains computationally intensive. Recent advancements include heuristics for variable ordering \cite{konrad2024FMCAD}, hybrid SCA-SAT solvers \cite{kaufmann2019verifying, Fujita24,chen2025reveal}, and addressing Gröbner basis limits through parallelization and memory optimization \cite{HPJHMHYB:2024}. 

Most current research employs static or dynamic ordering, often exploiting polarity. However, it remains unclear which methods suit specific circuit classes or if a universal, effective approach exists. This creates a vast configuration space of circuit types and optimizations. So far, no SCA-based tools exist to facilitate such exploration. Our tool TRACE exactly bridges this gap.

\section{SCA-based Proof Engine Architecture}\label{sec:proposed}
\begin{figure}[h!]
    \centering
    \includegraphics[width=\linewidth]{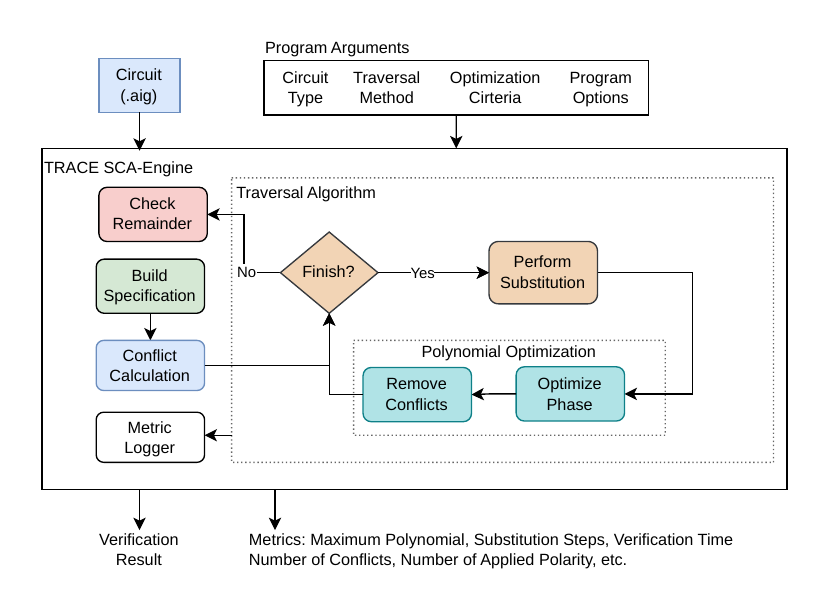}
    \caption{Overall TRACE Methodology}
    \label{fig:trace_methodology}
\end{figure}

The overall TRACE methodology is presented in Fig.~\ref{fig:trace_methodology}. The input to the tool is the AIG representation of the circuit. The tool also has several program arguments such as circuit type, traversal method and optimization criteria like polarity and conflict.
The verification process begins by constructing the specification polynomial. From there, signals within the polynomial are iteratively substituted based on the logic of the AIG nodes. A node becomes eligible for substitution once all successors of its output have been processed. This cycle continues until no further gates remain. If the resulting polynomial evaluates to zero, the verification is successful.
For the implementation, we utilize a gate-based substitution strategy. While grouping gates into larger functional blocks can streamline processing as proposed in~\cite{MGD:2022_RevSCA2}, identifying these blocks often introduces significant computational overhead. This is particularly challenging for circuits with optimized or unknown internal architectures.
For the substitution, several static orderings along with dynamic ordering methods are implemented. Also, for each of these methods, two optimization criteria—polarity and conflict are evaluated. After the verification process, a log file is generated which contains information like maximum polynomial, verification time, number of substitution steps, number of conflicts, number of polarities changed and many more. A detailed analysis of the supported methods is provided in the following section. 

\begin{figure}
    \centering
    \includegraphics[width=\linewidth]{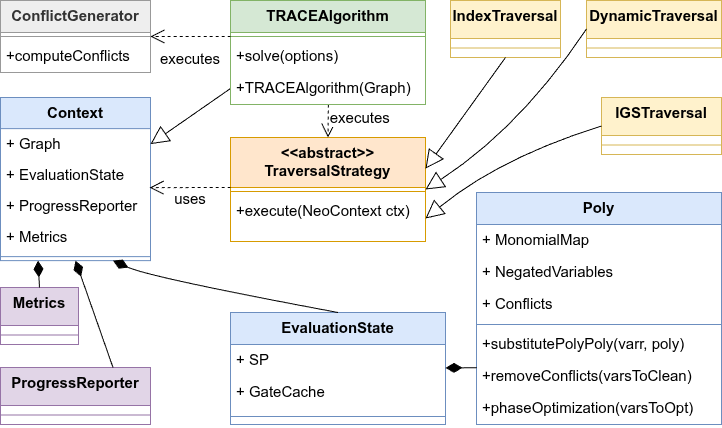}
    \caption{Software Architecture of the TRACE tool}
    \label{fig:arch}
\end{figure}

Fig.~\ref{fig:arch} presents the architecture of the tool as a UML diagram. Please note, that only the most relevant functions and members are presented. The entry point to the tool is the TRACEAlgorithm class, which is instantiated by providing the \emph{Design Under Verification} (DUV) as an AIG. To perform verification, the solve method is called. The main task executed by the TRACEAlgorithm is the setup of the verification context before starting the actual traversal. It sets up the specification polynomial and polynomial compression options, and it selects the correct traversal algorithm based on the provided options. If conflicts are used, they are calculated before the traversal. The Context is used to keep track of the state during the verification and share information between the TRACEAlgorithm and a traversal. It contains a Metrics object which tracks metrics during the verification, a ProgressReporter which enables the output of debug information for different phases of the verification, and the EvaluationState which contains the current SP as well as a cache for cache-based traversals. The SP uses the Poly class, which handles the efficient representation of a polynomial. It contains a set of monomials mapped to their coefficients, a set of currently negated nodes, as well as the calculated conflicts. The main functions are the substitution function, the conflict removal function, and the polarity optimization function. Different strategies for the traversal are implemented by inheriting from the abstract TraversalStrategy class.

While existing state-of-the-art tools are generally limited to verifying simple or complex multipliers, our framework offers the flexibility to explore a vast configuration space. This includes a diverse range of circuit types (adder, multiplier, MAC), traversal strategies, and optimization criteria, allowing for the study and more robust verification across varied architectures.

\section{Engine Capabilities}\label{sec:engine_capabilities}

\subsection{Specification Polynomial Generation}
For the following supported designs: ADD, MUL, and MAC with signed and unsigned option, TRACE generates the specification polynomial according to the provided input bit-width of the DUV. The specification polynomial is generated in two parts: a generic primary output section and a primary input section customized to the specific function being verified.
To manage the application of rules $r_1$ or $r_2$ for primary outputs, the node ID for the least significant primary output in the SP is set higher than any node ID in the DUV graph. These IDs are then substituted with the corresponding primary output IDs of the DUV before the main verification step begins.

For a MAC function $F -(A \times B) - S = 0$, where $F$ is the primary output vector,
and $A,B$ and $S$ are the primary input vectors the SP for a 2-bit design is defined in Eqn.~(\ref{eq:2mac}).
\begin{equation}\label{eq:2mac}
\begin{split}
SP_{MAC}(2) & :=
16F_4 + 8F_3 + 4F_2 + 2F_1 + F_0  \\
&- \big((2A_1 + A_0 ) \times (2B_1 + B_0)\big)\\
&- (8S_3 + 4S_2 + 2S_1 + S_0) = 0
\end{split}
\end{equation}
First, the primary output bit vector for $F$ is generated, with increasing IDs from least to most significant bit sizes. Next, the product of $A$ and $B$ is generated, for all pairwise bit combinations of $A_iB_j$, where each is scaled by the appropriate power of two coefficients $2^{i+j}$. Finally, the last part, the 3rd input for the accumulator is generated. Our tool also supports a polynomial generation (gen) mode where for any arbitrary design the polynomial expression of the circuit can be generated. Assuming that the primary output is encoded as a single integer.

\subsection{Polynomial Compression Techniques}
To ensure efficient verification using SCA, one must always overcome the challenge of intermediate polynomial blow-up. It has to be understood that, while important word level functions can be represented very compact, there are many functions that cannot be represented efficiently by SCA.
In the following, we describe two methods to modify the function a polynomial represents, reducing its size while maintaining verification correctness.
\subsubsection{Phase Optimization}
The first 
compression method is the phase optimization introduced in ~\cite{konrad2024FMCAD}. 
During verification each polynomial depends on a set of variables, which correspond to circuit signals. The polarity of those signals can easily be changed without losing any information about the function. Therefore, a signal $s$ is replaced with $1-s$ in the polynomial. When substituting signal $s$ later, this accordingly has to be taken into account. 
\begin{align}
    \label{eq:p1}& da+db+dc-dab-dac-dbc+dabc\\
    \label{eq:p2}=& d - d\overline{a} + d\overline{a}b + d\overline{a}c - d\overline{a}bc \\
    \label{eq:p3}=& d - d\overline{a}\overline{b} +d\overline{a}\overline{b}c\\
    \label{eq:p4}=& d - d\overline{a}\overline{b}\overline{c}
\end{align}
To illustrate the effect of phases on the polynomial size let us take a look at the polynomial from Eq.~\ref{eq:p1}. This polynomial is dependent on four variables and consists of 7 monomials. 
If only the variable $a$ in Eq.~\ref{eq:p2} is negated the monomial count is reduced to 5. After also negating $b$ and $c$ only two monomials remain in the end in Eq.~\ref{eq:p4}.
Negating $d$ in this case would lead to an increase in size. 
The challenge for an automated exploitation of this behavior is to determine which signals to negate.

There is no known method for precisely predicting the correct phases of signals, so in practice phases are determined by negating a signal and observing the effect on polynomial size.
This does not guarantee finding the optimal solution, phases can be considered to be a multidimensional optimization problem.
To reduce the complexity of the phase optimization, it was suggested in ~\cite{konrad2024FMCAD} to just optimize the phases of signals involved in the last substitution.
We follow this algorithm.
If $\mathcal{P}$ is a polynomial in which a signal $z$ was replaced using rule 3 (r3) $z = ab$ obtaining a new polynomial $\mathcal{P'}$, $a$ is negated in $\mathcal{P'}$. When this leads to a size reduction of $\mathcal{P'}$, $a$ is kept in negated form.
Afterward the same is done for signal $b$.

\subsubsection{Conflict Analysis}
Another important compression technique is \emph{conflict analysis} which was introduced in~\cite{KWDD:2026:arith} and in this paper we extend with an implication rule for zeros at AND outputs, as explained later. This optimization computes dependencies among internal signals as a preprocessing step and exploits this information during substitution to remove unnecessary monomials from intermediate polynomials.
To motivate this optimization, we first define the notion of a conflict and explain which monomials can be eliminated based on it.
Let the overall circuit represent a function $f:\mathcal{PI}\rightarrow\mathcal{PO}$ from the primary inputs $\mathcal{PI}$ to the primary outputs $\mathcal{PO}$.
During substitution, this function is conceptually decomposed into two parts: a function $g:\mathcal{PI}\rightarrow\mathcal{I}$ representing the remaining unsubstituted part of the circuit, and a function $h:\mathcal{I}\rightarrow\mathcal{PO}$ representing the already substituted part encoded by the current polynomial.
A key observation is that $g$ is not necessarily injective.
Hence, not every 
evaluation of the internal signals $\mathcal{I}$ is reachable from an assignment of the primary inputs.
In contrast, the polynomial representing $h$ is defined over all possible assignments to $\mathcal{I}$ and is therefore unaware of unreachable internal signal combinations.

In this setting, we call a pair of signals $(a,b)$ a \emph{conflict} if $a$ and $b$ cannot simultaneously assume the value 1.
Consequently, any monomial containing both $a$ and $b$ can be removed immediately, since it will always evaluate to 0 after further substitutions.
A well-known example arises in half adders: the sum and carry outputs can never be 1 at the same time ~\cite{MGD:2018b}.

\begin{figure}[]
     \centering
     \begin{subfigure}[b]{0.48\linewidth}
         \centering
         \includegraphics[width=\linewidth]{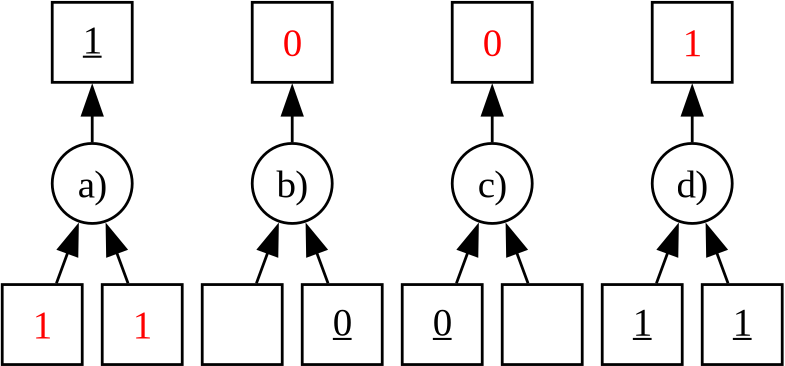}
         \caption{Implication rules used for conflict computation}
         \label{fig:conflict_implications}
     \end{subfigure}
     \hfill
     \begin{subfigure}[b]{0.48\linewidth}
         \centering
         \includegraphics[width=\linewidth]{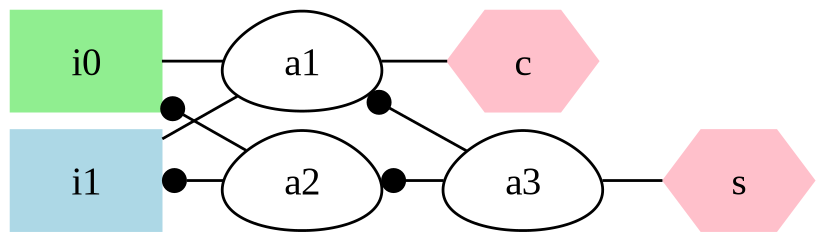}
         \caption{And-Inverter Graph of a half adder}
         \label{fig:ha_diagram}
     \end{subfigure}
        \caption{Implication rules and example circuit for conflict computation}
        \label{fig:combined_diagrams}
\end{figure}

To exploit conflicts for polynomial compaction, they must be derived automatically from the circuit graph.
We perform this computation as a preprocessing step before gate substitution.
The analysis is based on standard \emph{Automatic Test Pattern Generation} (ATPG) techniques.
For each signal, implications are computed independently for both assignments, i.e., assuming the signal to be 0 and assuming it to be 1.
For each assumption, implications are propagated through the graph using both forward and backward implication rules.

The implication rules used in our approach are shown in Fig.~\ref{fig:conflict_implications}.
Black underlined values denote the implication source, while red values indicate the implied assignments.
For example, a value 1 at the output of an AND gate implies that both inputs are 1, whereas a value 0 at one input implies that the output is 0.
Similarly, if both inputs are 1, the output must be 1.
For the case of a 0 at the output, we additionally check whether implication sets for both input assignments have already been computed.
If so, the corresponding implication sets are intersected, and the common assignments are treated as implied in this case as well.
Although this approximation is less precise than explicitly branching on all possible justifications, it is significantly cheaper.
Since implication analysis must be performed for every signal, full case splitting for each non-unique backward implication would be prohibitively expensive.
For AIGs, the implication rules must additionally account for edge negations; however, this extension is straightforward.

\begin{table}[]
    \centering
    \caption{Implications for the circuit in Fig.~\ref{fig:ha_diagram}}
    \label{tab:conflict_implications}
    \begin{tabular}{c|c||c|c|c|c|c}
        \toprule
         \textbf{Signals} & \textbf{Assumption} & \textbf{i0} & \textbf{i1} & \textbf{a1} & \textbf{a2} & \textbf{a3} \\
         \midrule
         i0 & 1/0 & 1/0 & x/x & x/0 & 0/x & x/x \\
         i1 & 1/0 & x/x & 1/0 & x/0 & 0/x & x/x \\
         a1 & 1/0 & 1/x & 1/x & 1/0 & \textbf{0}/x & 0/x \\
         a2 & 1/0 & 0/x & 0/x & 0/x & 1/0 & 0/x \\
         a3 & 1/0 & x/x & x/x & 0/x & 0/x & 1/0 \\
         \bottomrule
    \end{tabular}
\end{table}

Table~\ref{tab:conflict_implications} illustrates the implication analysis for the half-adder AIG shown in Fig.~\ref{fig:ha_diagram}.
Each row lists the implications obtained from one signal under both constant assignments (1/0).
Consider the row corresponding to the assumption $a1=1$.
In this case, $a3$ is directly implied to be 0, and both $i0$ and $i1$ are implied to be 1.
These implications further imply that $a2=0$.
Hence, $a1$ and $a2$ form a conflict.
Under every assignment that sets $a1$ to 1, $a2$ must be 0. Therefore, every monomial containing both $a1$ and $a2$ can be removed. To exploit the computed conflicts during substitution, the polynomial is checked after each substitution step for monomials containing conflicts involving the most recently substituted signals.
For efficiency, conflicts are stored in a hash map that associates each signal with the set of signals conflicting with it.

\begin{figure}[]
{\footnotesize
\begin{align*}
SP_0 \xrightarrow[s,c]{r1}
&= 2a1 + a3 && \\
SP_1 \xrightarrow[a3]{r5}
&= 2a1 + 1 -a1 -a2 + \mathcolor{red}{a1a2} &&= a1 + 1 - a2 \\
SP_2 \xrightarrow[a2]{r5}
&= a1 + 1 - 1 + i0 +i1 - i0i1 &&= a1 + i0 +i1 - i0i1\\
SP_3 \xrightarrow[a1]{r3}
&= i0i1 + i0 +i1 - i0i1 &&= i0 + i1
\end{align*}}
\caption{\normalsize Substitution steps for the circuit in Fig.~\ref{fig:ha_diagram} using conflict analysis}
\label{fig:conflict_substitition}
\end{figure}

Fig.~\ref{fig:conflict_substitition} demonstrates the substitution process with conflict removal for the half adder from Fig.~\ref{fig:ha_diagram}. We use the specification polynomial $2c+s$ which represents the arithmetic interpretation of a 1-bit adder.
In the first step, the outputs $c$ and $s$ are replaced by $a1$ and $a3$, respectively.
Next, $a3$ is substituted using rule~5 (r5).
This yields an intermediate polynomial containing the monomial $a1a2$, highlighted in red.
Since $a1$ and $a2$ are conflicting signals, this monomial can be removed immediately.
Subsequently, after substituting $a1$ and $a2$, the resulting polynomial is $i0+i1$, which matches the input specification of the adder.

Without conflict removal, substituting $a2$ in the monomial $a1a2$ would produce $a1-a1i0-a1i1+a1i0i1$.
This would substantially increase the size of $SP_2$.
For larger designs, eliminating such conflicting monomials can effectively prevent the explosion of intermediate polynomial size, as confirmed by our experimental results.

\subsection{Traversal Algorithms}
In the following, a selection of static and dynamic traversal algorithms is presented.

\subsubsection{Static Algorithms}
In the interest of space, only a selection of the supported static traversal algorithms is presented here: the \emph{Index-based} (IDX) method and \emph{Independent-Group-based Search} with its two variants, LSB-to-MSB ordering (IGS) and MSB-to-LSB ordering (IGSM). Other orderings like: \emph{Breath-First Search} (BFS), \emph{Depth-First Search} (DFS), \emph{Inductive Priority Simple} (IPS), \emph{Inductive Priority Complex} (IPC), IGS with Caching (IGSC), IPS with local BFS (IPSB) and IPS with local DFS search (IPSD) are also included in the tool.

The IDX ordering is generated by either Yosys or ABC synthesis, both of which provide an optimized dependency graph. This approach ensures a straightforward traversal from primary outputs to inputs, where each node ID in the graph is visited and substituted exactly once.
The second static ordering, IGS, is motivated by the need to prevent redundant computations. To avoid recomputing subgraphs repeatedly, each node is evaluated only once. To manage the complexity of verifying large graphs, the structure is partitioned into slices based on each primary output. This limits the number of nodes per traversal, thereby reducing the risk of intermediate term explosion. In this traversal algorithm, every time a node is visited, the incoming edge is logically removed; a node is only evaluated once all its incoming edges have been processed. This deferred evaluation of nodes with multiple dependencies ensures they are subject to only a single substitution, while simultaneously maintaining the graph in manageable, output-specific slices.

Considering the example of a 3-bit carry ripple adder from Fig.~\ref{fig:3rc}. The three orderings IDX, IGS and IGSM are presented below.
\begin{figure}[h!]
    \centering
    \includegraphics[width=0.8\linewidth]{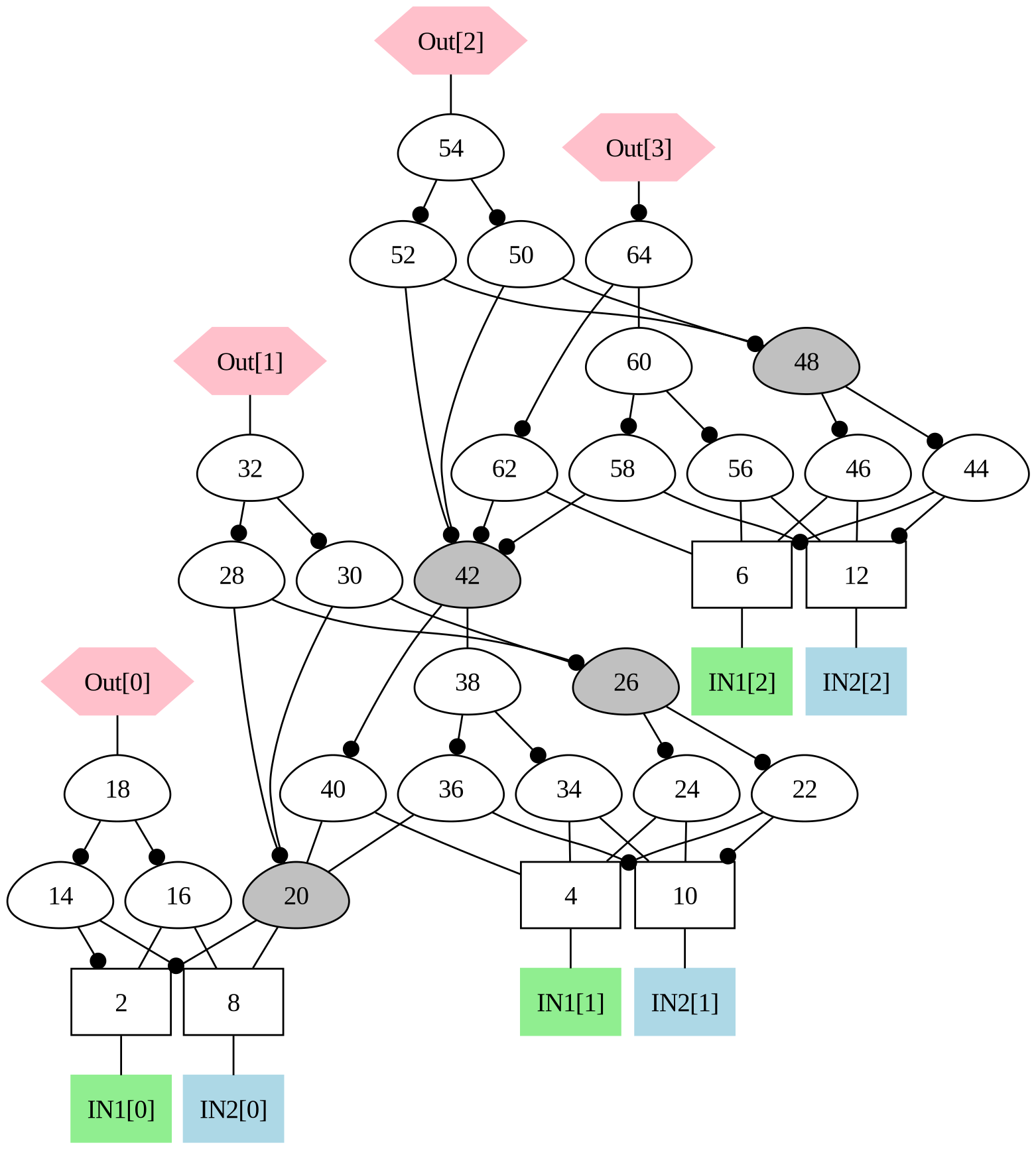}
    \caption{And-Inverter Graph of a 3-bit Carry Ripple Adder}
    \label{fig:3rc}
\end{figure}

\begin{description}
    \item[IDX]
    One execution sequence:
    (64, 62, 60, 58, 56, 54, 52, 50, 48, 46, 44, 42, 40,
    38, 36, 34, 32, 30, 28, 26, 24, 22, 20,
    18, 16, 14).
    \item[IGS]
    For each primary output we have one execution sequence (lsb to msb):
    Out[0]: (18, 16, 14),
    Out[1]: (32, 30, 28, 26, 24, 22),
    Out[2]: (54, 52, 50, 48, 46, 44),
    Out[3]: (64, 62, 60, 58, 56, 42, 40 38, 36, 34, 20).
    \item[IGSM]
    For each primary output we have one execution sequence (msb to lsb):
    Out[3]: (64, 62, 60, 58, 56),
    Out[2]: (54, 52, 50, 48, 46, 44, 42, 40, 38, 36),
    Out[1]: (32, 30, 28, 26, 24, 22, 20),
    Out[0]: (18, 16, 14).
\end{description}

The effectiveness of the traversal methods depends on the structure of the DUV graph. For this RC example, IDX and IGSM result in identical orderings, though IGSM groups the sequence by output. Conversely, IGS produces a different ordering that may lead to a less optimal solution than IGSM or IDX, depending on the DUV's internal dependencies.

\subsubsection{Dynamic Algorithms}
While the static algorithms calculate a substitution order based on graph structure alone, the dynamic traversal algorithm adapts the substitution order based on its effect on polynomial size. 
First introduced in ~\cite{MGS+:2020}, at each step one of the available substitutions is selected and performed. Afterward, the effect on the intermediate polynomial is measured as a growth factor. 
If the growth factor is above a specific threshold, the substitution is reversed and a different available substitution is executed.
This algorithm aims to traverse the circuit while avoiding cuts which are difficult to represent using an SCA polynomial. 
Considering the example of a 3-bit carry ripple adder from Fig. 7 the order produced by the dynamic traversal is presented below.
\begin{description}
    \item[DYN]
    One execution sequence:
    (64, 62, 54, 18, 16, 32, 30, 52, 50, 28, 14, 48,
    46, 44, 60, 56, 58, 42, 40, 26, 24, 22,
    38, 36, 34, 20).
\end{description}


\section{Experimental Evaluation}\label{sec:experimental}

The TRACE engine is implemented using C++ and all the experiments presented are performed on an AMD Ryzen 7 PRO 5850U with 32GB main memory. 
This tool enables the exploration of a vast configuration space covering various circuit types, traversal algorithms, and optimization techniques as mentioned in the previous section. Thereby, it allows designers to identify and select the most effective techniques for a specific circuit architecture.
The experiments were conducted on two sets of benchmark circuits: one generated using Genmul and Genmac and another generated using the Yosys\cite{yosys} and ABC\cite{abc:2018} tool. In principle, we can perform experiments using all supported traversal strategies and optimization factors like polarity and conflict; however, due to space limitations, a subset of traversal algorithms was selected with and without optimization factors.
For all experiments, a 60-minute timeout (T.O.) was set. Cases where 32GB was not sufficient to represent the intermediate polynomial are marked with memory timeout (M.T.O).

\subsection{Benchmark Generation}
In this section, the generation of the two main circuit types: structural and behavioral circuits are presented.

\subsubsection{Structural Circuit Generation}

Multipliers (MUL) are generated using Genmul and for Adders (ADD) and \emph{Multiply-and-Accumulate} (MAC) we use an in-house tool Genmac. Seven different adder types: \emph{Ripple Carry} (RC), \emph{Carry Look-Ahead} (CL), \emph{Ladner-Fischer} (LF), \emph{Kogge-Stone} (KS), \emph{Brent-Kung} (BK), \emph{Carry Skip} (CS), \emph{Serial Prefix} (SP) and four different multiplier types: Array (AR), \emph{Dadda Tree} (DT), \emph{Counter-based Wallace Tree} (CWT) and \emph{Wallace Tree} (WT) were considered. This resulted in 28 different multiplier configurations with seven final stage adders.
For the MAC generation the multiplier output must be added with another input of appropriate size. As the MUL consists of three stages: the 
\emph{Partial Product Generator} (PPG), the \emph{Partial Product Accumulator} (PPA), and the \emph{Final Stage Adder} (FSA), the third operand of MAC is integrated into the PPA stage rather than being added after the final product summation. By feeding both the PPG outputs and third operand directly into the PPA, the architecture eliminates the dedicated accumulation adder. This reduction in the critical path enhances the MAC unit’s speed and significantly optimizes power and area efficiency~\cite{swartzlander1980merged}. Finally, 28 different MAC architectures were designed. We categorize these as structural circuits because their underlying architectural design is explicitly known.

\subsubsection{Behavioral Circuit Generation}
Behavioral design involves defining hardware using high-level Verilog descriptions, which are subsequently processed by a synthesis tool. In this study, we utilized Yosys and ABC to transform the behavioral Verilog code into an AIG representation. Because the synthesis engine determines the final gate-level mapping and optimization strategies, the exact underlying architecture remains abstracted from the user. Circuits for both MUL and MAC are generated using this method.

\begin{table}[h!]
    \centering
    \caption{Longest path and nodes for 8-bit Multiplier and MAC designs}
    \label{tab:area_delay}
    \begin{tabular}{c|rr|rr}
    \toprule
         \textbf{Design} & \multicolumn{2}{c|}{\textbf{Multiplier}}  & \multicolumn{2}{c}{\textbf{MAC}}\\
         & Nodes &  Longest Path &  Nodes &  Longest Path\\\midrule
        $AR\circ RC$        &  1278 & 121 & 1666 & 94 \\
        $DT\circ LF$        &  1270 & 47  & 1706 & 48 \\
        $WT\circ KS$        &  1386 & 47  & 1825 & 49 \\\midrule
        Behavioral          & 1010  & 58  & 1324 & 58 \\
    \bottomrule
    \end{tabular}
\end{table}

Table~\ref{tab:area_delay} shows the structural designs occasionally outperform behavioral designs in terms of the longest path. However, for both the MAC and Multiplier units, the behavioral approach proves significantly more efficient, generating a more optimized design with a lower overall node count.

\subsection{Tool Performance}
Table~\ref{tab:features} shows the results for simple, moderate, and complex multiplier design ($AR\circ RC, DT\circ LF$ and $WT\circ KS$).
Three static ordering: IDX, IGS and IGSM and a dynamic ordering with no optimization (DYN), dynamic ordering with phase (DYN+P) and dynamic with phase and conflict (DYN+P+C) are evaluated for all multiplier designs from 8 to 64 bit. For all the methods Maximum Polynomial (MP) and Verification Time (VT) in seconds, which includes conflict analysis, polarity optimization and the verification itself, are reported.

\begin{table*}[t]
\centering
\caption{Structural Multiplier verification Capabilities for different Traversal Orderings}
\label{tab:features}
\begin{tabular}{ll||rr|rr|rr||rr|rr|rr}
\toprule
\textbf{Bit} & \textbf{Benchmark}
    & \multicolumn{6}{c||}{\textbf{Static Ordering}}
    & \multicolumn{6}{c}{\textbf{Dynamic Ordering}} \\
 &   &
\multicolumn{2}{c|}{IDX} &
\multicolumn{2}{c|}{IGS} &
\multicolumn{2}{c||}{IGSM} &
\multicolumn{2}{c|}{DYN} &
\multicolumn{2}{c|}{DYN+P} &
\multicolumn{2}{c}{DYN+P+C}\\
&
& MP & VT & MP & VT
& MP & VT & MP & VT
& MP & VT & MP & VT \\
\midrule
\multirow{3}{*}{8}
 & $AR\circ RC$ &  103& <0.01&   141 & 0.02&   107& 0.02&    143& 0.01&    125&0.02&   125 & 0.02\\
 & $DT\circ LF$ &  351& 0.01&   T.O.& T.O.&   232& 0.02&    791& 0.12&    327&0.21&   144 & 0.03\\
 & $WT\circ KS$ & T.O.& T.O.&   22309091& 381.39&  T.O.& T.O.&  20379&5.74&    339&0.03&   146 & 0.23\\
\midrule
\multirow{3}{*}{16}
 & $AR\circ RC$ &  311&0.03 & 413& 0.10& 315& 0.10&   415& 0.14&     402&  0.24& 397&0.23\\
 & $DT\circ LF$ &  667&0.04 & T.O.& T.O. & 819& 0.11&  3840&2.20&    5574&26.95& 524&0.38\\
 & $WT\circ KS$ & T.O.&T.O. & T.O.& T.O. & T.O.& T.O. &  T.O.& T.O.&    2197&16.49& 596&0.36\\
\midrule
\multirow{3}{*}{32}
 & $AR\circ RC$ & 1111& 0.32& 1341& 0.68& 1115& 0.62& 1344& 1.73& 1330& 3.25& 1330& 3.10\\
 & $DT\circ LF$ & 1575& 0.35& T.O.& T.O. & 3094& 0.69&  T.O.& T.O.  &  T.O.& T.O.  & 1769& 4.74\\
 & $WT\circ KS$ & T.O.& T.O.& T.O.& T.O. & T.O.& T.O. &  T.O.& T.O.  &  20165.& 1398.59   & 5208& 5.06\\
\midrule
\multirow{3}{*}{64}
 & $AR\circ RC$ & 4247& 47.11& 4733& 6.12& 4251& 6.02& 4736& 23.90& 4722& 54.80& 4722& 50.23\\
 & $DT\circ LF$ & 4819& 48.53& T.O.& T.O. & 13296& 7.03&  T.O.& T.O.  &  T.O.& T.O.  & 5739& 68.52\\
 & $WT\circ KS$ & T.O.& T.O.& T.O.& T.O. & T.O.& T.O. &  T.O.& T.O.  &  T.O.& T.O.  & 27310& 98.84\\
\bottomrule
\end{tabular}
\end{table*}

\begin{figure}[h!]
    \centering
    \includegraphics[width=0.85\linewidth]{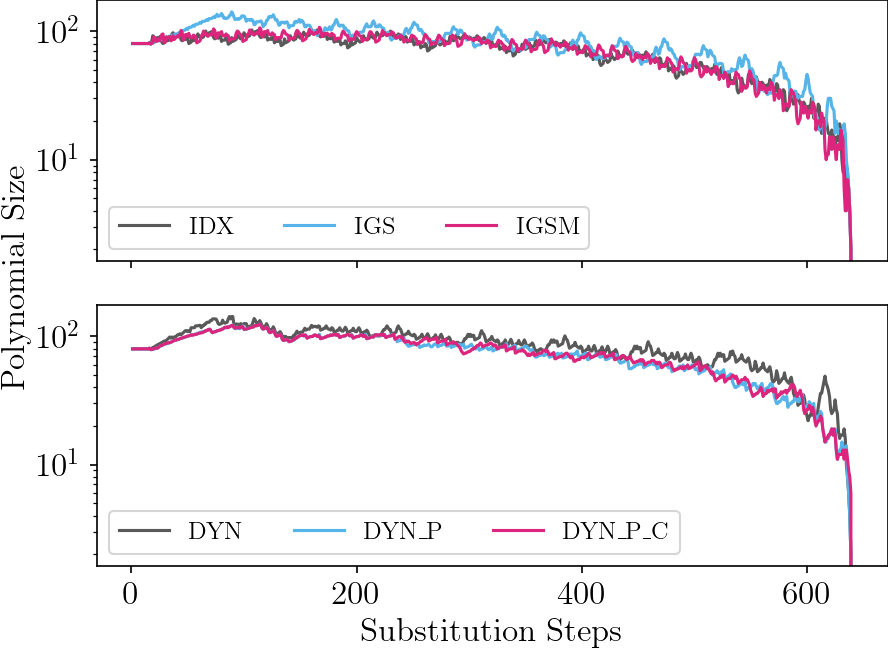}
    \caption{8-bit Simple Multiplier ($AR\circ RC$)}
    \label{fig:mul_ar_rc_8}
\end{figure}

\begin{figure}[h!]
    \centering
    \includegraphics[width=0.85\linewidth]{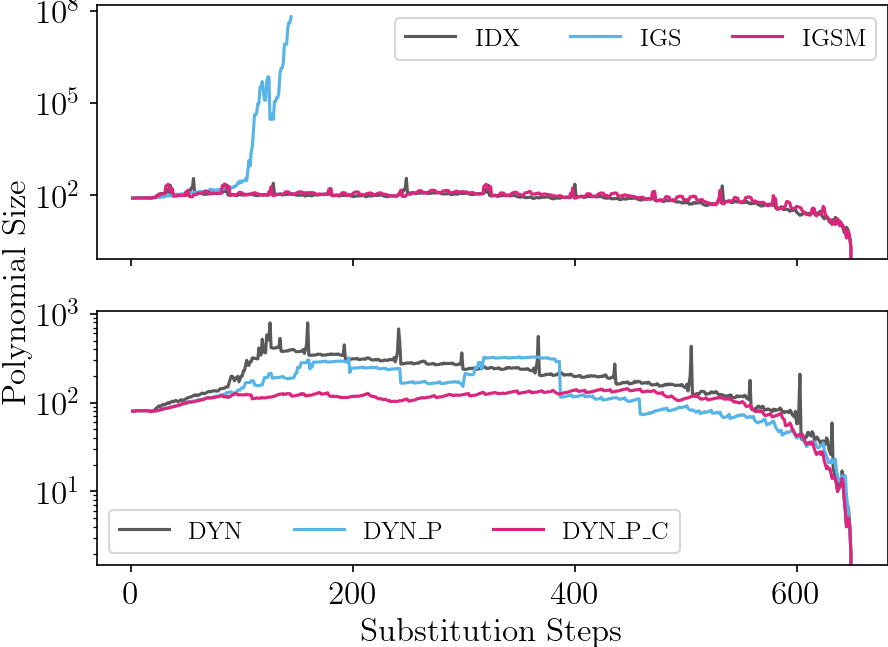}
    \caption{8-bit Moderately Complex Multiplier ($DT\circ LF$)}
    \label{fig:mul_dt_lf_8}
\end{figure}

To illustrate the influence of different traversal algorithms and optimizations on different multiplier architectures, we present two figures: Fig.~\ref{fig:mul_ar_rc_8} and Fig.~\ref{fig:mul_dt_lf_8}. The polynomial size is on the y-axis, and the substitution step are on the x-axis. The plots clearly show that for simple and moderately complex designs, simple static ordering methods perform very well. See IDX or IGSM for $AR \circ RC$ or $DT \circ LF$; these perform better than both DYN and DYN+P.


\begin{figure}[h!]
    \centering
    \includegraphics[width=0.85\linewidth]{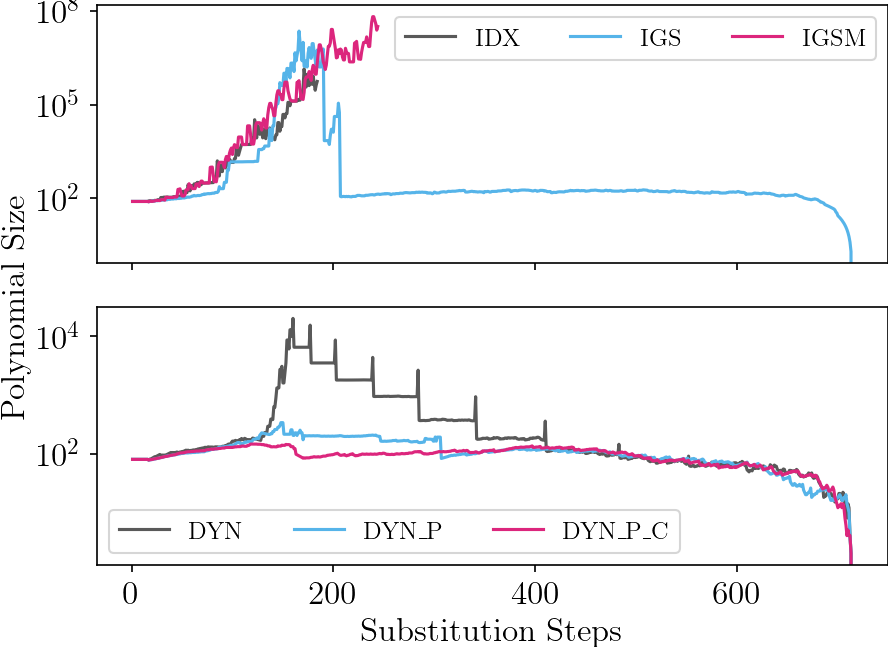}
    \caption{8-bit Complex Multiplier ($WT\circ KS$)}
    \label{fig:mul_wt_ks_8}
\end{figure}

Fig.~\ref{fig:mul_wt_ks_8} shows the performance of an 8-bit complex multiplier ($WT\circ KS$). From the plot, it is clear that for complex structures, static ordering often fails to verify the circuit within the time limit, whereas dynamic ordering with phase and conflicts is found to be most effective. This is also true for larger circuits (see Table~\ref{tab:features}). It is also interesting to see that our tool can handle very large polynomial sizes ($\sim$22 million) in verifying $WT\circ KS$ using IGS. These results clearly show the power of the TRACE tool in exploring the entire configuration space. This suggests that no single verification methodology is universally optimal across all circuit architectures.

\subsection{Structural Benchmark Performance}

\begin{table}[htpb]
    \centering
     \caption{Results for Structural MUL and MAC }
     {\setlength{\tabcolsep}{4pt}
    \begin{tabular}{r|rrrr|rrrr}
    \toprule
     \textbf{Method} &
     \multicolumn{4}{c|}{\textbf{MUL}} &
     \multicolumn{4}{c}{\textbf{MAC}} \\
     &
    8 & 16 & 32 & 64 &
    8 & 16 & 32 & 64\\\midrule
    TRACE       & 28 & 27 &\textbf{ 27} & 24 & 27 & 27 & 24 & 22\\ 
    RevSCA~\cite{MGD:2022_RevSCA2}/Ext~\cite{WDD:2025:fdl}  & 28 & 28 & 28 & 28 & 28 & 12 & 12 & 12\\ 
    AMulet2.2~\cite{kaufmann2021amulet}   & 23 & 8 & 8 & 8 & - & - & - & -\\ 
    DynPhaseOrderOpt~\cite{konrad2024FMCAD}   & 28 &  28 & 28 & 23 & - & - & - & -\\ 
    \bottomrule
    \end{tabular}
    }
    \label{tab:structural_benchmark}
\end{table}

The structural benchmarks were generated using Genmul and Genmac as mentioned in Section 5.1. Using four PPAs and seven FSAs we have a total of 28 benchmarks for both MUL and MAC. For conciseness, here we provide an overview of the results. Full experimental data for all configurations can be generated using our executable tool.
Table~\ref{tab:structural_benchmark} presents the overall results for MUL and MAC designs, comparing TRACE with RevSCA, AMulet, DynPhaseOrderOpt and RevSCA extension. TRACE was evaluated using DYN+P+C, while the other engines do not support configurable strategies. This lack of configurability was one of the main motivations for creating TRACE. While RevSCA successfully verified all multiplier designs (8 to 64-bit), TRACE encountered limitations at higher bit-widths, specifically failing on $CWT\circ CS$ at 16-bit, $DT\circ CS$ at 32-bit and four complex architectures (e.g., $DT\circ CS$, $CWT\circ BK$, $CWT\circ CS$, $CWT\circ SE$) at 64-bit. DynPhaseOrderOpt can verify one more circuit compared to TRACE across all multipliers, whereas AMulet shows the lowest verification coverage, verifying approximately 25\% of the overall circuits from 16-bit onwards.
Whereas for MAC designs, the RevSCA extension verified fewer than half of the cases from 16-bit onwards, TRACE successfully verified the majority of designs for 8, 16, 32 and 64-bit widths. 
AMulet and DynPhaseOrderOpt do not support MAC verification.

While RevSCA is better suited for multiplier designs, it fails to verify the majority of MAC designs. In contrast, TRACE successfully verifies most 8- and 16-bit configurations and significantly outperforms the RevSCA extension on MAC designs overall.


\subsection{Behavioral Benchmark Performance}
\begin{table*}[ht]
\centering
\caption{Behavioral MAC Verification Results for Different Substitution Orderings and Optimizations}
\label{tab:behavioural_results}
{\setlength{\tabcolsep}{4pt}\scriptsize
\begin{tabular}{rr|rr|rr||rr|rrr|rrrr||rr|rr}
\toprule
\textbf{B} & \textbf{NN} &
 \multicolumn{2}{c|}{\textbf{IDX}}  &
 \multicolumn{2}{c||}{\textbf{IGSM}}  &
 \multicolumn{2}{c|}{\textbf{Dynamic}} &
 \multicolumn{3}{c|}{\textbf{Dynamic+Phase}}  &
 \multicolumn{4}{c||}{\textbf{Dynamic+Phase+Conflict}}  &
 \multicolumn{2}{c|}{\textbf{RevSCAExt}~\cite{WDD:2025:fdl}} &
 \multicolumn{2}{c}{\textbf{Transformation}}\\
 & &
MP & VT &
MP & VT & 
MP & VT & 
MP & VT & PN &
MP & VT & PN & NC &
MP & VT & MP & VT \\
\midrule
2 &   38&       30&  <0.001&       35&  <0.01&     36&  <0.01&      26&  <0.01&   17&       22& <0.01&    7&    12 &      19&    0.001&      13&  0.001 \\
3 &   81&       60&   0.001&       64&  <0.01&     58&  <0.01&      37&  <0.01&   41&       34& <0.01&   16&    29 &      29&    0.002&      23&  0.002 \\
4 &  147&      650&   0.002&     1001&   0.01&     74&  <0.01&      64&   0.01&   73&       50& <0.01&   29&    53 &      49&    0.003&      34&  0.002 \\\midrule
5 &  237&    42777&   0.147&    35042&   0.12&    103&  <0.01&      93&   0.01&  120&       67& <0.01&   48&    85 &    4297&    0.115&      50&  0.006 \\
6 &  334&    41741&   0.098&    66524&   0.27&    192&   0.01&     142&   0.02&  169&       84&  0.01&   65&   119 &  418960&  176.248&      90&  0.009 \\
7 &  451&  1385673&  10.251&  2507453&  28.48&    233&   0.02&     194&   0.02&  225&      109&  0.01&   84&   162 &  732122&  274.251&     146&  0.009 \\
8 &  596& 21398915& 270.854& 12758712& 305.30&    376&   0.03&     287&   0.08&  300&      130&  0.02&  110&   218 & 5819510& 1445.410&     336&  0.033 \\\midrule
9 &  751&   M.T.O.&  M.T.O.&   M.T.O.& M.T.O.&    338&   0.05&     371&   0.15&  383&      158&  0.03&  160&   271 & 5888912& 2472.290&  M.T.O.& M.T.O. \\
10&  919&   M.T.O.&  M.T.O.&   M.T.O.& M.T.O.&    398&   0.04&     478&   0.47&  475&      183&  0.04&  175&   337 &  M.T.O.&   M.T.O.&    1261&  0.128 \\
11& 1106&   M.T.O.&  M.T.O.&   M.T.O.& M.T.O.&    393&   0.08&     445&   0.13&  548&      215&  0.05&  213&   403 &  M.T.O.&   M.T.O.&   50908& 15.624 \\
12& 1312&   M.T.O.&  M.T.O.&   M.T.O.& M.T.O.&    996&   0.24&     728&   0.66&  666&      238&  0.07&  232&   483 &  M.T.O.&   M.T.O.&  M.T.O.& M.T.O. \\\midrule
13& 1536&   M.T.O.&  M.T.O.&   M.T.O.& M.T.O.&    782&   0.20&     678&   0.45&  764&      282&  0.09&  301&   565 &  M.T.O.&   M.T.O.&  M.T.O.& M.T.O. \\
14& 1776&   M.T.O.&  M.T.O.&   M.T.O.& M.T.O.&    736&   0.19&     848&   0.98&  893&      339&  0.12&  278&   666 &  M.T.O.&   M.T.O.&  M.T.O.& M.T.O. \\
15& 2019&   M.T.O.&  M.T.O.&   M.T.O.& M.T.O.&    918&   0.15&    1135&   1.49&  992&      373&  0.14&  311&   746 &  M.T.O.&   M.T.O.&  M.T.O.& M.T.O. \\\midrule\midrule
16& 2019&   M.T.O.&  M.T.O.&   M.T.O.& M.T.O.&   1196&   0.44&    1315&   1.99& 1152&      412&  0.19&  451&    851&  M.T.O.&   M.T.O.&  M.T.O.& M.T.O. \\
32& 2019&   M.T.O.&  M.T.O.&   M.T.O.& M.T.O.&   6138&   7.36&   14943& 278.03& 4426&     2574&  2.86& 1726&   8368&  M.T.O.&   M.T.O.&  M.T.O.& M.T.O. \\
64& 2019&   M.T.O.&  M.T.O.&   M.T.O.& M.T.O.&  27731& 219.38&    T.O.&   T.O.& T.O.&    15135& 86.06& 7279& 265698&  M.T.O.&   M.T.O.&  M.T.O.& M.T.O. \\
\bottomrule
\end{tabular}
}
\end{table*}

\begin{figure}[h!]
    \centering
    \includegraphics[width=0.9\linewidth]{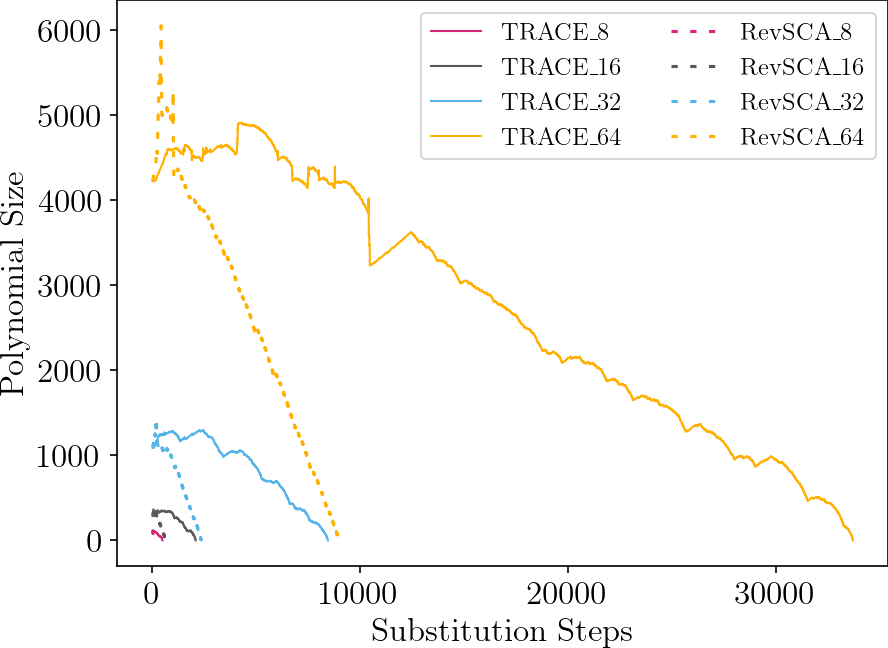}
    \caption{Behavioral Multiplier design verification}
    \label{fig:mul_behavioral_trace_revsca}
\end{figure}

In this subsection, we analyze the results of behavioral multiplier and MAC designs, comparing them against RevSCA and its extension. Fig.~\ref{fig:mul_behavioral_trace_revsca} illustrates the performance of 8, 16, 32, and 64-bit behavioral multipliers for both TRACE and RevSCA. While both methods demonstrate high performance, TRACE achieves a smaller maximum polynomial. However, RevSCA requires significantly fewer substitution steps because it utilizes atomic block-based substitution, whereas TRACE relies on gate-based substitution.

Behavioral MAC designs, on the other hand, have posed significant verification challenges. Table~\ref{tab:behavioural_results} compares our  approaches against existing work~\cite{WDD:2025:fdl,WDD:2025:dvcon}. From left to right, the table is organized as follows: the first two columns specify the bit-width (B) and the number of AIG nodes (NN), followed by groups for substitution order and optimization strategies. These categories—IDX, IGSM, Dynamic, Dynamic+Phase, and Dynamic+Phase+Conflict—are compared against extensions of RevSCA for MAC and Transformation-based methods. For each method, a subset of the following metrics is reported: Maximum Polynomial (MP) size, Verification Time (VT), number of Polarity Negations (PN), and the number of Conflicts (NC). Previously, the state-of-the-art method for behavioral MAC could verify only up to 11 bits using a transformation approach~\cite{WDD:2025:dvcon} based on extended RevSCA. In comparison, without transformation it reaches its limit at 9-bits.

The results clearly demonstrate the limitations of static ordering for behavioral or complex MACs; using IDX and IGSM, only circuits up to 8-bits can be verified. While the MP reaches 21 million for IDX and 5.8 million for RevSCA extensions, DYN+P+C reaches an MP of only 130. This clearly shows that for complex designs with unknown internal structures, only our dynamic ordering combined with polarity and conflict optimization is capable of meeting the verification requirements. Experiments also demonstrate the power of our tool, which is capable of verifying circuits up to 64 bits; in contrast, both static methods and state-of-the-art extensions of RevSCA can only verify circuits up to 11-bits. These results emphasize that dynamic ordering and advanced techniques like phase and conflict identification are vital for addressing the intermediate polynomial expansion issues typically encountered in SCA-based verification. Our tool shows the effectiveness of static ordering for simple designs, while highlighting the necessity of phase and conflict optimizations for verifying complex systems.
\section{Conclusion}\label{sec:conclusion}

TRACE tool provides a highly efficient environment, allowing for the experimental comparison of various traversal strategies and optimizations to determine their impact on memory usage and verification time across different arithmetic architectures. It extends the verification horizon of SCA beyond multipliers.  
Experimental evaluations demonstrate that while static ordering methods are effective for simple to medium-complexity designs, they consistently fail to complete verification for highly complex architectures. In contrast, dynamic ordering provides a more robust solution for complex designs; however, without additional optimization, the maximum polynomial size can still grow beyond manageable limits. Our results show that when standard static orderings and state-of-the-art methods fail, the integration of polarity and conflict-based optimizations within TRACE becomes essential to maintaining a compact symbolic representation. Finally, as a general framework, it delivers superior results across a diverse range of arithmetic circuits while providing the flexibility to explore a large configuration space, ensuring both scalability and high-performance verification compared to existing SCA tools.




\newpage
\bibliographystyle{ACM-Reference-Format}
\bibliography{bib/lit_header_short, bib/lit_my_long, bib/lit_others, bib/lit_pfv_koselleck, bib/lit_focus, bib/lit_agra, bib/lit_sca}

@inproceedings{MGD:2018b,
  author    = {Alireza Mahzoon and Daniel Gro{\ss}e and Rolf Drechsler},
  title     = {{PolyCleaner:} Clean your Polynomials before Backward Rewriting to Verify Million-gate Multipliers},
  booktitle = iccad,
  pages     = {129:1--129:8},
  year      = {2018},
  mycomment     = {{\bf (Best Paper Award)}}
}

@InProceedings{SGK+:2016,
  author        = {Amr Sayed-Ahmed and Daniel Gro{\ss}e and Ulrich K\"uhne and Mathias Soeken and Rolf Drechsler},
  title		= {Formal Verification of Integer Multipliers by Combining {Gr\"obner} Basis with Logic Reduction},
  booktitle     = date,
  year		= 2016,
  pages         = {1048--1053},
  mycomment     = {{\bf (Best Paper Candidate)}}
}

@inproceedings{moskewicz2001chaff,
  title={{Chaff: Engineering an efficient SAT solver}},
  author={Moskewicz, Matthew W and Madigan, Conor F and Zhao, Ying and Zhang, Lintao and Malik, Sharad},
  booktitle={Proceedings of the 38th annual Design Automation Conference},
  pages={530--535},
  year={2001}
}

@ARTICLE{YBL+:2016,
   LANGUAGE  = {USenglish},
   AUTHOR    = {Cunxi Yu and Walter Brown and Duo Liu and Andre Rossi and Maciej Ciesielski},
   TITLE     = {Formal verification of arithmetic circuits by function extraction},
   JOURNAL   = TCAD,
   YEAR      = 2016,
   VOLUME    = 35,
   NUMBER    = 12,
   PAGES     = {2131-2142},
}

@ARTICLE{YCM:2017,
   LANGUAGE  = {USenglish},
   AUTHOR    = {Cunxi Yu and Maciej Ciesielski and Alan Mishchenko},
   TITLE     = {Fast Algebraic Rewriting Based on And-Inverter Graphs},
   JOURNAL   = TCAD,
   YEAR      = 2017,
   VOLUME    = 37,
   NUMBER    = 9,
   PAGES     = {1907-1911},
}

@ARTICLE{FA:2015,
   LANGUAGE  = {USenglish},
   AUTHOR    = {Farimah Farahmandi and Bijan Alizadeh},
   TITLE     = {Gr\"obner basis based formal verification of large arithmetic circuits using Gaussian elimination and cone-based polynomial extraction},
   JOURNAL   = micmic,
   YEAR      = 2015,
   VOLUME    = 39,
   NUMBER    = 2,
   PAGES     = {83-96},
}

@MISC{yosys,
    author = {Claire Xenia Wolf},
    title = {{Yosys Open SYnthesis Suite}},
    howpublished = "\url{https://yosyshq.net/yosys/}",  year={2024}
}

@inproceedings{RBK:2017,
 author = {Daniela Ritirc and Armin Biere and Manuel Kauers},
 title = {Column-Wise Verification of Multipliers Using Computer Algebra},
 booktitle = fmcad,
 pages     = {23--30},
 year = {2017}
}

@inproceedings{RBK:2018,
	author    = {Daniela Ritirc and Armin Biere and Manuel Kauers},
	title     = {Improving and extending the algebraic approach for verifying gate-level multipliers},
	booktitle = date,
	pages     = {1556--1561},
	year      = {2018},
}

@misc{abc:2018,
  title = {ABC: A System for Sequential Synthesis and Verification},
  howpublished = "available at \url{https://people.eecs.berkeley.edu/~alanmi/abc/}",
  year = {2018}, 
}

@article{KBK:2019,
  author    = {Daniela Kaufmann and Armin Biere and Manuel Kauers},
  title     = {{Incremental column-wise verification of arithmetic circuits using computer algebra}},
  journal   = fm,
  month = {Feb.},
  year      = {2019},
}

@inproceedings{MGS+:2020,
  author    = {Alireza Mahzoon and
               Daniel Gro{\ss}e and
               Christoph Scholl and
               Rolf Drechsler},
  title     = {Towards Formal Verification of Optimized and Industrial Multipliers},
  booktitle = date,
  pages     = {544--549},
  year      = {2020}
}

@article{B:1986,
  author    = {Randal E. Bryant},
  title     = {{Graph-Based Algorithms for Boolean Function Manipulation}},
  journal   = toc,
  volume    = {35},
  number    = {8},
  pages     = {677--691},
  year      = {1986}
}

@inproceedings{kaufmann2019verifying,
  title={{Verifying large multipliers by combining SAT and computer algebra}},
  author={Kaufmann, Daniela and Biere, Armin and Kauers, Manuel},
  booktitle={2019 Formal Methods in Computer Aided Design (FMCAD)},
  pages={28--36},
  year={2019},
  organization={IEEE}
}

@article{swartzlander1980merged,
  title={Merged arithmetic},
  author={Swartzlander},
  journal={IEEE Transactions on Computers},
  volume={100},
  number={10},
  pages={946--950},
  year={1980},
  publisher={IEEE}
}

@STRING{toc	= {TC} }

@STRING{tcad	= {TCAD} }

@STRING{fm	= {Formal Methods in System Design: An International
		  Journal} }

@STRING{micmic  = {MICPRO }}

@STRING{arith   = {ARITH} }

@STRING{iccad	= {ICCAD} }

@STRING{date	= {DATE} }

@STRING{fdl	= {FDL} }

@STRING{fmcad	= {FMCAD} }

@STRING{dvcon_europe   = {DVCon Europe} }

@STRING{sat     = {Theory and Applications of Satisfiability Testing} }

@inproceedings{WDD:2025:date,
    author = "Lennart Weingarten and Kamalika Datta and Rolf Drechsler",
    title = {{Late Breaking Results: Towards Efficient Formal Verification of Dot Product Architectures}},
    booktitle = date,
    year = {2025},
    numpages = {},
    pages={1-2},
    publisher = {{IEEE}},
    address = {},
    doi={10.23919/DATE64628.2025.10992946}
}

@inproceedings{WDD:2025:fdl,
    author={Lennart Weingarten and Kamalika Datta and Rolf Drechsler},
    title = {{ForMAt: Formal Verification of Scalable Multiply and Accumulate Unit}},
    booktitle = fdl,
    pages={1-7},
    year = 2025
}

@inproceedings{WDD:2025:dvcon,
    author={Lennart Weingarten and Kamalika Datta and Rolf Drechsler},
    title = {{Transformation-Aided Verification of MAC Designs using Symbolic Computer Algebra}},
    journal = dvcon_europe,
    year = 2025,
    pages={1-7}
}

@inproceedings{KWDD:2026:arith,
    author={Jan Kleinekath{\"o}fer and Lennart Weingarten and Kamalika Datta and Rolf Drechsler},
    title = {{A Conflict-Aware Learning Approach to SCA Verification for MAC Architectures}},
    booktitle = arith,
    year = 2026,
    pages={1-8}
}

@ARTICLE{Fujita24,
  author={Li, Rui and Li, Lin and Yu, Heng and Fujita, Masahiro and Jiang, Weixiong and Ha, Yajun},
  journal={IEEE Transactions on Computer-Aided Design of Integrated Circuits and Systems}, 
  title={{RefSCAT: Formal Verification of Logic-Optimized Multipliers via Automated Reference Multiplier Generation and SCA-SAT Synergy}}, 
  year={2024},
  volume={},
  number={},
  pages={1-1},
  doi={10.1109/TCAD.2024.3442987}
}

@INPROCEEDINGS{HPJHMHYB:2024,
  author={Liu, Hongduo and Liao, Peiyu and Huang, Junhua and Zhen, Hui-Ling and Yuan, Mingxuan and Ho, Tsung-Yi and Yu, Bei},
  booktitle=date, 
  title={{Parallel Gröbner Basis Rewriting and Memory Optimization for Efficient Multiplier Verification}}, 
  year={2024},
  volume={},
  number={},
  pages={1-6},
  doi={10.23919/DATE58400.2024.10546568}
}

@inproceedings{konrad2024FMCAD,
  title={{Symbolic Computer Algebra for Multipliers Revisited-It's All About Orders and Phases}},
  author={Konrad, Alexander and Scholl, Christoph},
  booktitle={2024 Formal Methods in Computer-Aided Design (FMCAD)},
  year={2024},
}

@ARTICLE{MGD:2022_RevSCA2,
  author={Mahzoon, Alireza and Große, Daniel and Drechsler, Rolf},
  journal={IEEE Transactions on Computer-Aided Design of Integrated Circuits and Systems}, 
  title={{RevSCA-2.0: SCA-Based Formal Verification of Nontrivial Multipliers Using Reverse Engineering and Local Vanishing Removal}}, 
  year={2022},
  volume={41},
  number={5},
  pages={1573-1586},
  doi={10.1109/TCAD.2021.3083682}}

@article{chen2025reveal,
  title={{ReVEAL: GNN-Guided Reverse Engineering for Formal Verification of Optimized Multipliers}},
  author={Chen, Chen and Kaufmann, Daniela and Deng, Chenhui and Song, Zhan and Zhang, Hongce and Yu, Cunxi},
  journal={arXiv preprint arXiv:2512.22260},
  year={2025}
}

@inproceedings{kaufmann2021amulet,
  title={{AMulet 2.0 for verifying multiplier circuits}},
  author={Kaufmann, Daniela and Biere, Armin},
  booktitle={International Conference on Tools and Algorithms for the Construction and Analysis of Systems},
  pages={357--364},
  year={2021},
  organization={Springer}
}

@article{biere2007aiger,
  title={{The AIGER and-inverter graph (AIG) format version 20071012}},
  author={Biere, Armin},
  year={2007}
}

\end{document}